\documentclass[]{IEEEtran}

\usepackage{graphicx}
\usepackage{url}
\usepackage[utf8]{inputenc}
\usepackage{amsmath}
\usepackage{amsfonts}
\usepackage{amssymb}
\usepackage{upgreek}
\DeclareRobustCommand*{\IEEEauthorrefmark}[1]{%
  \raisebox{0pt}[0pt][0pt]{\textsuperscript{\footnotesize\ensuremath{#1}}}}

\begin{document}

%\title{(Title in 24-point Times font)}
% If the \LARGE is deleted, the title font defaults to  24-point.
% Actually, 
% the \LARGE sets the title at 17 pt, which is close enough to 18-point.
%+++++++++++++++++++++++++++++++++++++++++++
\title{\LARGE High-resolution hydrogen spectroscopy  \\ and the Proton Radius Puzzle}
%\author{\authorblockN{J. Clerk Maxwell}
%\authorblockA{School of Electrical and\\Computer Engineering\\
%Somewhere Institute of Technology\\
%City, State 54321--0000\\
%Email: maxwell@curl.edu}
%\and
%\authorblockN{Michael Faraday}
%\authorblockA{(List authors on this line using 12 point Times font\\ - use a second line if necessary)\\
%Microwave Research\\
%City, State/Region, Mail/Zip Code, Country\\
%\and
%\authorblockN{Andr\'e M. Amp\`ere \\ }
%\authorblockA{Starfleet Academy\\
%San Francisco, CA 96678-2391\\

%+++++++++++++++++++++++++++++++++++++++++++++++++++

% avoiding spaces at the end of the author lines is not a problem with
% conference papers because we don't use \thanks or \IEEEmembership

% Another example.
\author{\IEEEauthorblockN{Simon Thomas\IEEEauthorrefmark{1},
H\'el\`ene Fleurbaey\IEEEauthorrefmark{1}\IEEEauthorrefmark{*}, Sandrine Galtier\IEEEauthorrefmark{1}\IEEEauthorrefmark{\dagger}, \\ 
Lucile Julien\IEEEauthorrefmark{1}, Fran\c cois Biraben\IEEEauthorrefmark{1} and Fran\c cois Nez\IEEEauthorrefmark{1}} \bigskip \\
\IEEEauthorblockA{\IEEEauthorrefmark{1}Laboratoire Kastler Brossel, Sorbonne Universit\'e, CNRS, ENS - Universit\'e PSL, Coll\`ege de France, \\ 
4 place Jussieu, case 74, 75005 Paris, France.} \\
\IEEEauthorblockA{\IEEEauthorrefmark{*}{\it Present address}: National Institute of Standards and Technology, \\
100 Bureau Drive, Gaithersburg, MD 20899, USA.} \\
\IEEEauthorblockA{\IEEEauthorrefmark{\dagger}{\it Present address}: Institut Lumi\`ere Mati\`ere, UMR 5306, Universit\'e Lyon 1-CNRS, Universit\'e de Lyon, \\
69622 Villeurbanne {\sc cedex}, France.} \\
\textit{Corresponding author: simon.thomas@lkb.upmc.fr.}}
 %\\
%\IEEEauthorblockA{\IEEEauthorrefmark{2}Affiliation2}\\(Please list e-mail address of corresponding author; other e-mails are optional)}
%\specialpapernotice{(Invited Paper)}
\maketitle

\begin{abstract}
High resolution spectroscopy of the hydrogen atom takes on particular importance in the new SI, as it allows to accurately determine fundamental constants, such as the Rydberg constant and the proton charge radius. 
Recently, the second most precisely measured transition frequency in hydrogen, $1S-3S$, was obtained in our group. 
In the context of the Proton Radius Puzzle, this result calls for further investigation.
\end{abstract}
%\IEEEoverridecommandlockouts
%\begin{IEEEkeywords}
%Hydrogen Spectroscopy, Proton Charge Radius, Rydberg Constant.
%\end{IEEEkeywords}

% for peerreview papers, inserts a page break and creates the second title.
% Will be ignored for other modes.
\IEEEpeerreviewmaketitle

% Added this command to remove (gobble!) the page numbers (the correct pages will be added by the IEEE editors, if at all). GG October 2015.
\pagenumbering{gobble}

\section{Introduction}
\subsection{Determining $R_\infty$ and $r_p$}
As the simplest atomic system, the hydrogen atom can be described with great accuracy by the theory of Quantum Electrodynamics (QED). More precisely, the fine structure of its energy levels can be calculated as a function of a reduced number of fundamental constants as:
\begin{equation}
E_{njl}=-\frac{hcR_{\infty}}{1+\frac{m_e}{m_p}} \left[\frac{1}{n^2}+F_{njl}\left(\alpha,\frac{m_e}{m_p},\frac{m_e}{m_\mu}\right)-\delta_{l0}\frac{C_{NS}}{n^3}r^2_p\right]
\end{equation}
where $R_{\infty}$ is the Rydberg constant, $\alpha$ the fine structure constant, ${m_e}/{m_p}$ and ${m_e}/{m_\mu}$ the electron-to-proton and electron-to-muon mass ratios and $r_p^2$ the second moment of the charge distribution of the proton. At first order, the nuclear size contribution is, for S states, $C_{NS}=\frac{4}{3} (4\pi)^2 R^2_{\infty}(1+\frac{m_e}{m_p})^{-2}\alpha^{-2}$.

Assuming QED to be correct, values of those fundamental constants can be deduced from the comparison between the theory and high resolution spectroscopy measurements.
As $\alpha$ and the mass ratios can be measured by other means with a sufficient accuracy, only the determination of $R_\infty$ and $r_p$ is actually critically dependent from such a comparison. In practice, two transition frequencies are thus needed to jointly extract values of these two constants.

Reversely, assessing the consistency of the different values of $R_\infty$ and $r_p$ obtained from the spectroscopy of different transitions therefore provides a test of QED.

\subsection{The Proton Radius Puzzle}
A disagreement among determinations of the proton radius $r_p$ was first noticed in 2010 \cite{nature2010,aldoscience}. 
The CREMA collaboration, then achieving the laser spectroscopy of muonic hydrogen, measured a value of $r_p$ that was ten times more precise, but also 4 \% smaller than the previously accepted value (Figure 1).
The corresponding discrepancy is $5.6\sigma$, as compared to the CODATA-2014 recommended value, which aggregates all precise former measurements in hydrogen spectroscopy and electron-to-proton scattering \cite{codata2014}.

This disagreement, also known as the \emph{Proton Radius Puzzle}, has since then stimulated an intense research activity \cite{carlson2015}.
Recently, two hydrogen spectroscopy results have notably been published: the $2S-4P$ transition frequency, measured at MPQ \cite{beyer2017}, and the $1S-3S$ transition frequency that we measured at LKB \cite{fleurbaey18}.
When combined with the precisely measured $1S-2S$ transition frequency \cite{parthey}, the first one yields a value of the proton radius in agreement with the muonic hydrogen value, whereas the second one is consistent with the CODATA-2014 value. 
As the disagreement persists, the hypothesis of an unsuspected systematic effect having affected hydrogen spectroscopy measurements cannot be discarded.

\begin{figure}
\includegraphics[width=\columnwidth]{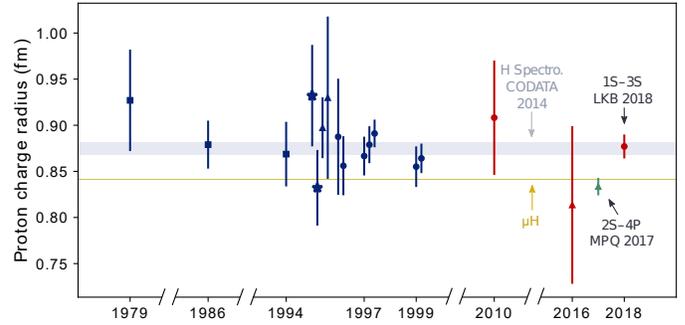}%
\caption{\label{PRP}%\col 
Proton charge radius values from H spectroscopy, with 1$\sigma$ errorbars. 
Squares are RF measurements of the $2S-2P$ transition, achieved in Harvard and University of Sussex. Combinations of the $1S-2S$ transition with other optical transitions obtained at Yale, MPQ and LKB are represented with stars, triangles and circles, respectively.
The hydrogen spectroscopy measurements included in the CODATA-2014 adjustment are in blue; their average corresponds to the light blue bar. The $1S-3S$ transition appears in red. The thin golden bar is the result from muonic hydrogen spectroscopy.
}
\end{figure}

\subsection{Spectroscopy of the $1S-3S$ transition}
The $1S-3S$ transition frequency of the hydrogen atom has been measured in our group since the late nineties \cite{bourzeix96, hagel2002}. 
It is now the second most precisely studied transition in hydrogen, after the $1S-2S$ transition.
Forbidden as a dipolar electric transition, it can be excited by two counter-propagating photons, therefore without Doppler broadening.
This allows us to almost access the natural linewidth of the transition, $\varGamma$ $\approx 1$ MHz, and to determine its frequency with an uncertainty of a few kHz (below $10^{-12}$ in relative uncertainty).

Should the Proton Radius Puzzle be reformulated as the search for a potential systematic effect, the spectroscopy of the $1S-3S$ transition is to play a crucial role in this search, 
as the only ongoing experiment in agreement with the formerly obtained values of $R_\infty$ and $r_p$.
In this regard, after a brief review of the principle of our experiment, we will present in what follows the current status of our work.

\section{Principle of the experiment}
\subsection{Experimental setup}

\begin{figure*}
\includegraphics[width=\textwidth]{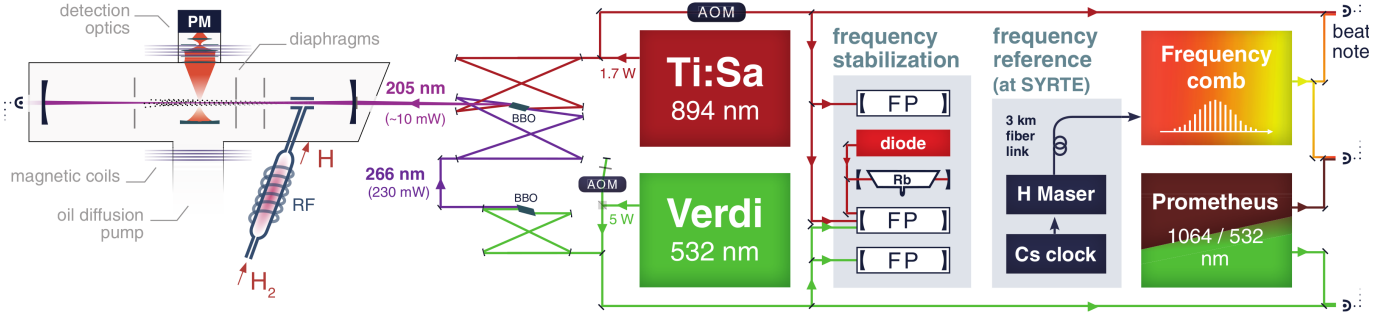}%
\caption{\label{Manip}%\col 
Simplified view of the experimental setup.}
\end{figure*}

Figure \ref{Manip} displays a schematic overview of the experiment conducted in Paris.
An effusive beam of atomic hydrogen is produced at room temperature by a radiofrequency (RF) discharge and directed, through a nozzle, colinearly with a laser beam at 205 nm. Propagating in a Fabry-Perot cavity under vacuum, this laser beam undergoes a frequency scan, performed by an acousto-optic modulator (AOM), in order to excite the $1S_{1/2}^{F=1}-3S_{1/2}^{F=1}$ transition of the atoms. The resonance is then observed by collecting, in a photomultiplier(PM), the photons at 656 nm emitted by the fluorescence from the $3S$ level to the $2P$ level. 
Figure \ref{Fig_spectres} shows an example of the recorded signal. 

The laser beam at 205 nm is obtained by sum frequency generation in a BBO cristal, between a TiSa laser at 894 nm and a frequency-doubled Verdi laser at 532 nm. This unique scheme provides between 10 and $15\ \text{mW}$ of cw light at 205 nm \cite{sandrineoptcom}. The frequency stability of those lasers is ensured by successive locks to stable Fabry-Perot cavities (FP) and to a two-photon transition of Rubidium \cite{rubidium}.
This locking scheme feeds a double-pass AOM which provides an additional phase stabilization to the 532-nm laser beam.

An optical frequency comb is used to measure the frequencies of the two laser sources at 532 nm and 894 nm. It is referenced to a hydrogen maser, whose frequency is continuously monitored relatively to the Cs clock at the LNE-SYRTE, Observatoire de Paris \cite{epjd1998}.
A frequency-doubled Nd:YAG laser emitting both at 1064 and 532 nm (Prometheus) is used as a transfer laser between our 532-nm laser source and the frequency comb, optimized in the infrared range.

\subsection{Systematic effects}
Although cancelled at first order, thanks to the counter-propagating configuration, 
the Doppler shift on the measured transition frequency is at second order not negligible:
\begin{equation}
\Delta_{SOD}= -\frac{v^2}{2c^2}\nu_L
\end{equation}
with $v$ the atomic velocity and $\nu_L$ the laser frequency.
It is our main systematic effect: for hydrogen atoms with an average velocity of 3 km/s, the second-order Doppler shift (SOD) amounts to about -135 kHz,
that is eighty times larger than the currently-achieved uncertainty. 
It is corrected from our experimental spectra by fitting them with a theoretical lineshape, which is integrated over the velocity distribution of our atomic beam. 
Section (3) describes this process.

Two other smaller effects, a light shift and a pressure shift, are corrected by experimental extrapolation to respectively zero laser power and zero hydrogen pressure. They typically amount to less than 10 kHz. 
Eventually, the measured transition frequency is corrected so as to relate to the French \emph{mise en pratique} of the second.

\section{Second-order Doppler shift}
\subsection{Motional Stark shift method}
\begin{figure}
  \includegraphics[width=\columnwidth]{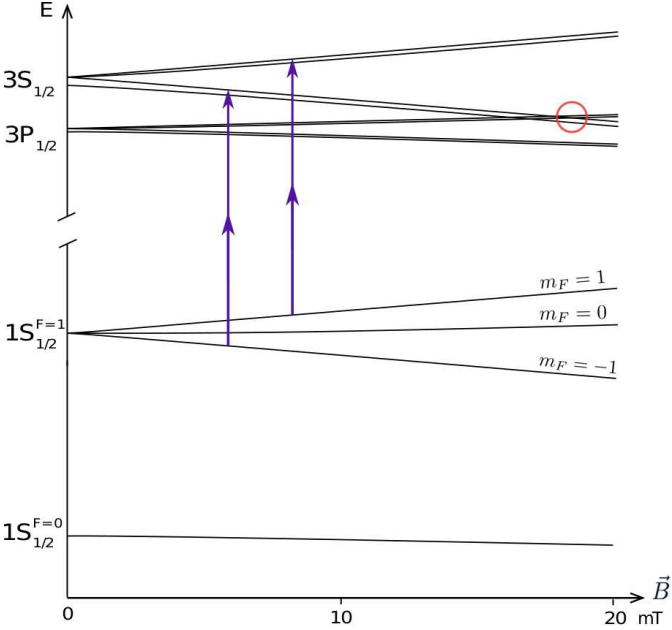}%
  \caption{\label{Zeeman}%\col 
Zeeman diagram of the energy levels of interest of the hydrogen atom.
Encircled is the anti-crossing of which the motional Stark shift method takes advantage.
Conservation of angular momentum implies that the transitions driven by two counter-propagating photons couple states of same $(F,m_F)$.
The $(F=1,m_F=0)$ subtransition is sensitive to the Zeeman effect: the measurement of its frequency allows to calibrate the value of the applied $\vec{B}$ field \cite{gaetthese}.
The $(F=1,m_F=\pm 1)$ subtransitions (double arrows) are at first order not Zeeman-shifted. Because of their natural linewidth, they cannot be resolved. They are the ones we measure to determine the atomic velocity distribution.
}
\end{figure}

The correction of the SOD requires the determination of the atomic velocity distribution within our hydrogen beam.
To this end, we apply a magnetic field $\vec{B}$ perpendicular to the direction of the beam \cite{gaetPRL},\cite{biraben91}. The motional electric field perceived by the atoms induces a quadratic Stark effect, that shifts their energy levels depending on their velocity $\vec{v}$:
\begin{equation}
\Delta_{Stark} \propto |\vec{v}\times\vec{B}|^2
\end{equation}
This shift is magnified at the vicinity of an anticrossing between energy levels coupled by the Stark effect. The Zeeman effect leads to such an anticrossing, between the $3S_{1/2}^{F=1,m_F=-1}$ and the $3P_{1/2}^{F=1,m_F=0}$ levels, at $B=18$ mT (Figure \ref{Zeeman}). 

The variation of this motional Stark shift with the applied magnetic field carries information on the velocity of the atoms. It is thus possible to adjust the parameters of a theoretical velocity distribution, by fitting experimental spectra obtained at different $\vec{B}$ values. 

\subsection{Theoretical lineshape}
The theoretical lineshape, with which the experimental spectra are fitted, is classically obtained by calculating the fluorescence probability of the hydrogen atom as a function of the laser frequency $\nu_L=\omega_L/2\pi$ \cite{arnoult2010}.
The hydrogen atom is here considered as having a velocity $\vec{v}$ in the laboratory frame, experiencing an homogeneous magnetic field $\vec{B}$ perpendicular to $\vec{v}$. 
It can be described by its density matrix, that verifies the Optical Bloch Equation:
\begin{equation}
\frac{d\rho}{d t}=\frac{i}{\hbar}\left[\rho,\hat{H}_0+\hat{H}_{S}+\hat{H}_{Z}+\hat{H}_{2\gamma}\right]+\left\{\frac{d\rho}{d t}\right\}_{rel}
\label{Bloch}
\end{equation}

The considered hamiltonian can be decomposed as a sum of four terms.
First, $\hat{H}_0$ describes the fine and hyperfine structure of the unperturbed atom, as a function of the unknown $1S-3S$ centroid transition frequency $\nu_{1S-3S}$. 
Second, the motional Stark effect is accounted for by 
$\hat{H}_S=-q\vec{r}\cdot(\vec{v}\times\vec{B})$
where $\vec{r}$ is the position operator and $q$ the charge of the electron.
Third, Zeeman and diamagnetic effects are described by:
\begin{equation}
\hat{H}_Z=-\frac{q\vec{B}}{2}\cdot\left(\frac{\vec{L}}{\mu}+\frac{g_s\vec{S}}{m_e}-\frac{g_N\vec{I}}{m_p}\right) + \frac{(q\vec{r}\times\vec{B})^2}{8\mu}
\end{equation} 
where $\mu$ is the reduced mass of the atomic system, and $g_s$ and $g_N$ are the Land\'e g-factor of the bound electron and of the proton. 
With respect to the free electron g-factor $g_e$, a relativistic correction is applied, depending on the principal quantum number $n$ of the involved state: $g_s=g_e(1-\alpha^2/3n^2)$ \cite{BetheSalpeter}.
And fourth, the two-photon transition hamiltonian is, in the rotating wave approximation and taking into account the SOD:
\begin{equation}
\hat{H}_{2\gamma}=\frac{\hbar\Omega_{eg}}{2} e^{-i\omega_L(2+\frac{v^2}{c^2}) t}|e\rangle\langle g| + \frac{\hbar\Omega_{eg}}{2} e^{i\omega_L(2+\frac{v^2}{c^2}) t}|g\rangle\langle e|
\end{equation} 
where $|g\rangle$ and $|e\rangle$ are the ground and excited states of the atom between which the two-photon selection rule is verified ($\Delta F=0, \Delta m_F=0$).
The two-photon Rabi frequency can be expressed in the dipole approximation as \cite{biraben73}:
\begin{equation}
\Omega_{eg} = \frac{q^2E^2}{\hbar^2}\sum_{k}\frac{\langle e|\vec{r}\cdot\vec{\epsilon}|k\rangle\langle k|\vec{r}\cdot\vec{\epsilon}|g\rangle}{\omega_L-\omega_{kg}}
\end{equation}
where $E$ and $\epsilon$ are the amplitude and polarization of the laser field. In our case, $\epsilon$ is colinear to the direction of the applied magnetic field, taken as our quantization axis. $\omega_{kg}$ denotes the angular transition frequency between the levels $|g\rangle$ and $|k\rangle$.
We will consider $\Omega$ to be independent of $\omega_L$, in the absence of energy levels at resonance with the laser frequency. 

In our case, we are only interested in calculating the density matrix coefficients corresponding to energy levels of principal quantum number $n=3$. With this restriction, the relaxation due to spontaneous emission, as described by the Lindblad operator, becomes:
\begin{equation}
\left\{\frac{d\rho_{ij}}{d t}\right\}_{rel}=-\frac{\varGamma_i+\varGamma_j}{2}\rho_{ij}
\end{equation}
with $\varGamma_i$ the decay rate of the level $|i\rangle$.

In the rotating wave approximation, equation (\ref{Bloch}) yields a system of equations with time-independent coefficients, that can be directly solved in the stationary regime.

Once the stationary state of the atom is known, and neglecting in first approximation quantum interference effects, the fluorescence probability per steradian and per unit of time can be calculated as: 
\begin{equation}
F_{fluo} = \frac{\alpha}{2\pi c^2}\sum_{f,\vec{\epsilon}}\sum_{i}\omega_{if}^3|\langle f|\vec{r}\cdot\vec{\epsilon}|i\rangle|^2\rho_{ii}
\end{equation}
Here, $i$ and $f$ denote the initial and final states between which radiative decay can occur and which lead to the emission of a photon with a wavelength and a polarization $\vec{\epsilon}$ detectable by our apparatus. Again in first approximation, we will only consider here the photons emitted along the axis of our photomultiplier. These approximations are further discussed below.

Eventually, this fluorescence is convoluted with a function describing an additional broadening, and integrated over the velocity distribution. A $1/v$ factor accounts for the lower excitation probability of the atoms experiencing a smaller interaction time with the laser beam:
\begin{equation}
F_{B,\sigma,v_0} = \int{\frac{\mathrm{d}v}{v}f_{\sigma,v_0}(v)F_{fluo}(\nu_L;\nu_{1S-3S},v,B)\ast F_{\varGamma}(\nu_L)}
\end{equation}

A global treatment of the line broadening is performed by employing an ad-hoc expression for the broadening function $F_{\varGamma}$.
Following ref. \cite{biraben79}, finite transit time induces a double-exponential shaped broadening, while collisional processes are essentially described by a lorentzian shape. As the latter accurately fits the broadening we experimentally observe, it is the one we used in our analysis.
%But as our experimental broadening results are well fitted by a lorentzian broadening, this latter function 
A more general function, such as a Voigt or pseudo-Voigt function, can as well be chosen, so as to treat in a less model-dependent manner potential sources of broadening. 
It was verified that such a choice does not change the final result by more than 100 Hz, and that the gaussian component of the best fitting pseudo-Voigt function (sum of a gaussian and a lorentzian profile) is at most 10 \%.

\subsection{Quantum interference effect}
A more complete calculation of the fluorescence probability involves interference terms, that induce an asymmetry of the lineshape, depending on the angle of emission of the fluorescence photon \cite{hessels10,hessels11}.
In the case of the $1S-3S$ transition, this effect is weak: integrated over our detection geometry, at zero magnetic field, it amounts to 0.6 kHz \cite{fleurbaey2017}.
It is thus simply included in our theoretical lineshape by shifting the considered value of $\nu_{1S-3S}$ by +0.6(2) kHz.

\subsection{Fit of a theoretical velocity distribution}
The atomic velocity distribution of the hydrogen beam can be modeled as a thermal effusive beam, with correction factors accounting for collisional processes \cite{JPCRD15}. A function $P$, parameterized with the adequate Knudsen number, detailed in \cite{olander}, describes the effect of interactions occuring within the nozzle, while an exponential decay is used to model an additional depletion of slow atoms \cite{champenois}:
\begin{equation}
f_{\sigma,v_0}(v) \propto v^{3}e^{-\frac{v^2}{2\sigma^2}}P(v/\sigma)e^{-\frac{v_0}{v}}
\end{equation}

Such correction factors improve the characterization of the velocity distribution, reducing by 3 to 6 \% the $\chi^2$ of the fit to all experimental data. 
Nevertheless, the parameter values of these correction factors have a limited influence on the final result. 
For instance, in order to see a variation of 1 kHz of $\nu_{1S-3S}$ fitted using $f_{\sigma,v_0}$, a variation of the value of $v_0$ by 8 \% would be required, and of the Knudsen number by more than an order of magnitude, while $\sigma$ would need to vary by 0.5 \%.

To determine the values of $\sigma$ and $v_0$, all experimental spectra are fitted with a function $c_1 F_{B,\sigma,v_0}(\nu_L;\nu_{1S-3S},\varGamma)+c_2$, where $\nu_{1S-3S}$, $\varGamma$, $c_1$ and $c_2$ are left floating, and $(\sigma,v_0)$ are taken from a grid of $N$ pairs of values.
For a given set of spectra, and given values of $\sigma$ and $v_0$, the $\chi^2$ of the values of $\nu_{1S-3S}$ obtained for the different spectra is calculated.
The resulting surface of $\chi ^2(\sigma,v_0)$ can then be fitted with a polynomial. Optimal values $(\sigma^{opt},v_0^{opt})$ are obtained at the minimum $\chi^2_{min}$ of this fitted surface. Their uncertainties are defined so that values of $\sigma$ and $v_0$ lying within the uncertainty range verify $\chi^2(\sigma,v_0)\leq \chi^2_{min}+\text{Max}(1,R_B^2)$, with the Birge ratio defined as $R_B^2=\chi^2_{min}/(N-2)$ \cite{theseHelene}. 

Once the velocity distribution is known, each experimental spectrum can be again fitted with the theoretical lineshape to yield its corresponding value of $\nu_{1S-3S}$. The uncertainty on $\nu_{1S-3S}$ is evaluated in the same manner as for $\sigma$ or $v_0$.

\section{Data analysis}
\subsection{Data sets}
Two measurement campaigns have been conducted, in 2013 and 2016-2017. 
More than 2700 experimental spectra of the $1S-3S$, $F=1$ transition have been acquired, each one corresponding to a typical integration time of 10 s per frequency point.
Details on these recordings can be found in \cite{fleurbaey18, theseHelene}. 
They were performed for various values of hydrogen pressure, laser power and applied magnetic field, in order to evaluate the aforementioned systematic effects.

The motional Stark shift method was carried out for four different subsets of spectra.
Those subsets, thereafter denoted by a,b,c and d, were respectively recorded in 2013 at $P=7.5\times 10^{-5}$ mbar (a); in 2016 and 2017 at $P=2.7\times 10^{-5}$ mbar (b and c) and at $P=2.0\times 10^{-4}$ mbar (d). 

\begin{figure}
  \includegraphics[width=230pt]{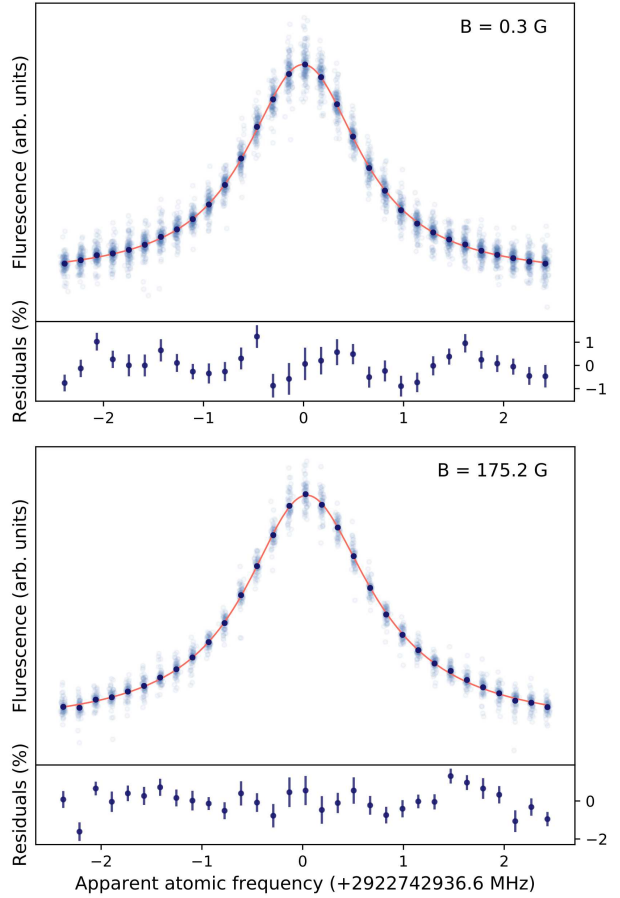}%
  \caption{\label{Fig_spectres}%\col 
Average of the experimental spectra of the subset (c), obtained at $B=0.3$ G (124 spectra) and $B=175.2$ G (61 spectra). For each magnetic field, the average spectrum is fitted with the theoretical lineshape described section 3.2 (red line); the residuals are shown below. Errorbars correspond to the standard deviation of the experimental points that are averaged at each frequency point. These experimental points, integrated each over 10 s, are shown in transparency. 
}
\end{figure}

\begin{figure}
  \includegraphics[width=230pt]{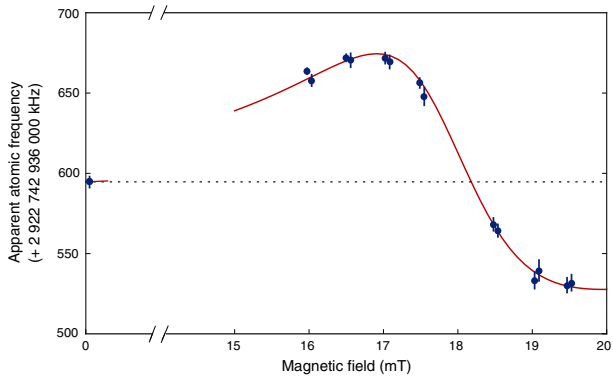}%
  \caption{\label{dispersion}%\col 
Experimental (dots) and calculated (line) apparent frequency of the $1S-3S$ $(F=1,m_F=\pm 1)$ transition, as a function of the applied magnetic field $\vec{B}$.
The experimental points correspond to the subset of spectra (c) (see below) fitted by a simple Lorentzian lineshape.
The velocity distribution parameters used in the calculation were deduced from the subset (c) with the fitting procedure presented in section 3.4.}
\end{figure}

\begin{figure*}
  \includegraphics[width=\textwidth]{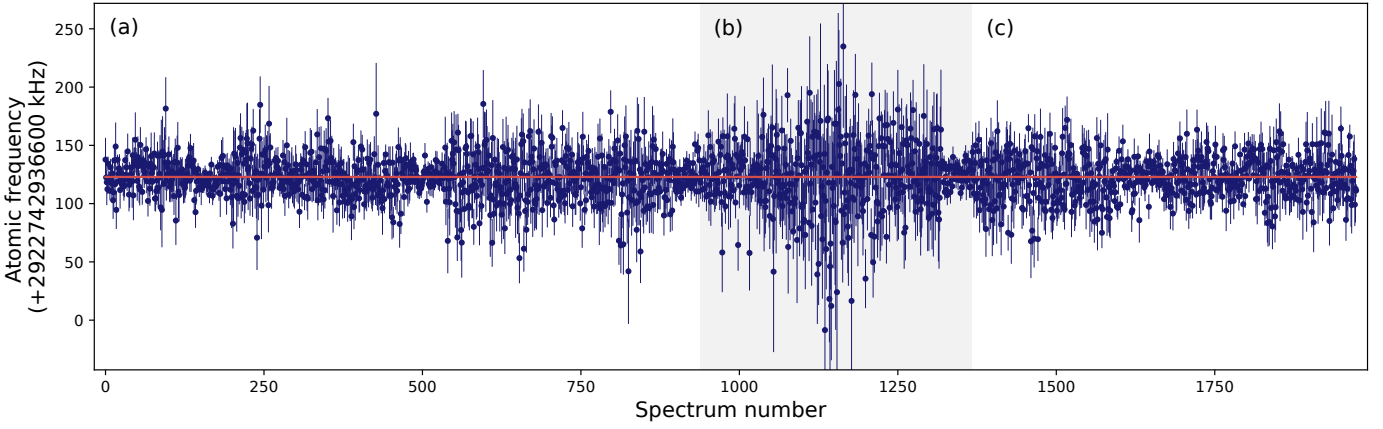}%
  \caption{\label{allspectra}%\col 
Fitted frequencies $\nu_{1S-3S}$ from all spectra of subsets (a), (b) and (c) (see text), after systematic effects correction, in chronological order of acquisition. 
Each value results from the fit of 31 frequency points, integrated each over 10 s. 
Among those spectra, 452 were recorded at $B=0.3$ G, and 1522 between 160 and 195 G: no $B$-dependent systematic shift could be uncovered.
The weighted average of all 1974 values is represented by the red line; the corresponding standard deviation is 0.29 kHz, the $\chi^2$ is 2223.4 and the Birge ratio is 1.06. 
For the sake of readability, spectra recorded at a higher hydrogen pressure are not shown here, as the associated uncertainties are much larger.
}
\end{figure*}

\begin{figure}
  \includegraphics[width=230pt]{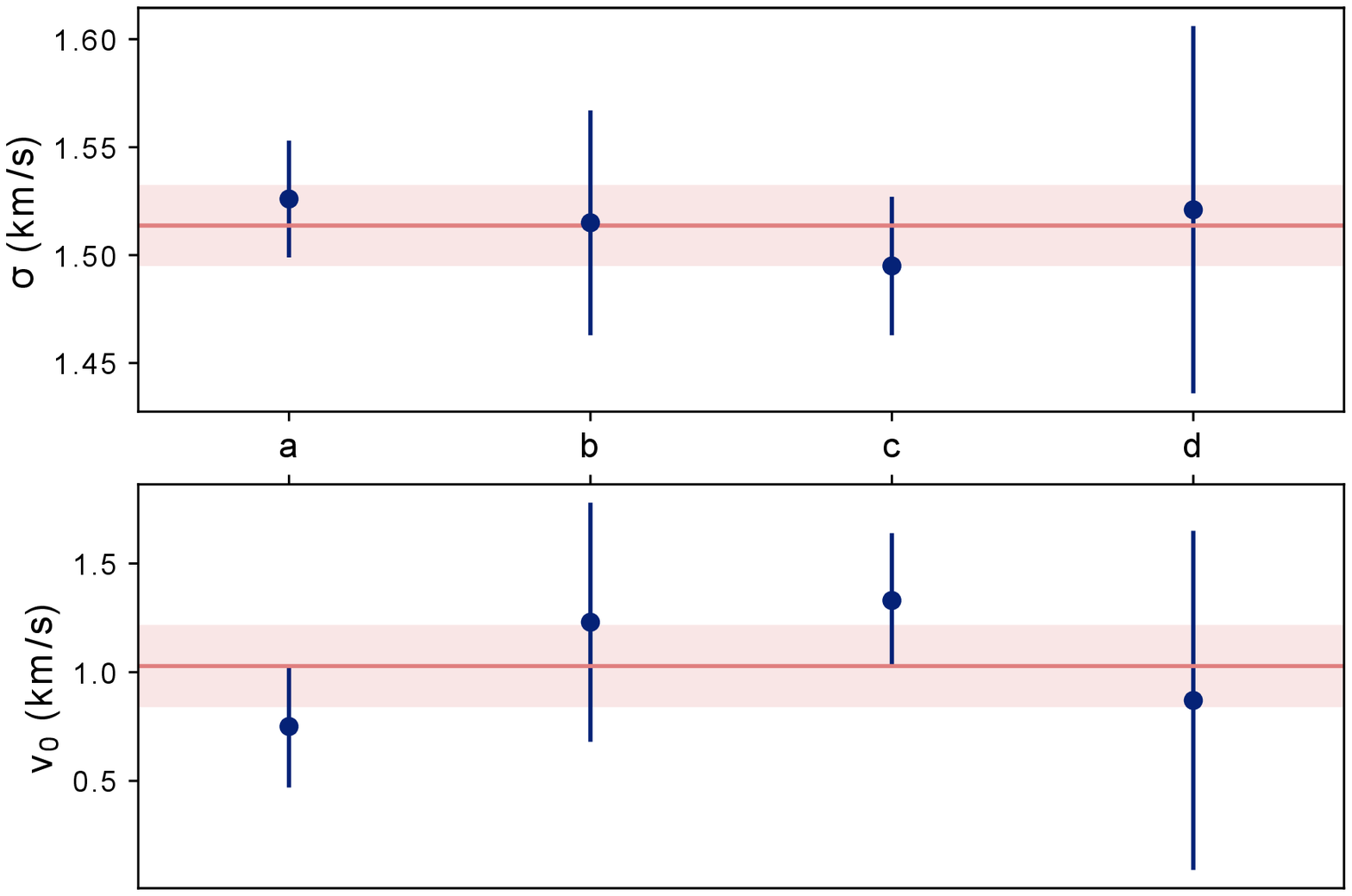}%
  \caption{\label{distribvitesses}%\col 
Values of the velocity distribution parameters $\sigma$ and $v_0$ determined at different hydrogen pressure values. 
The $\chi^2$ of the fit yielding such values is 974 for subset (a) (with 936 degrees of freedom); 422 for subset (b) (426 d.o.f.); 821 for subset (c) (606 d.o.f.) and 445 for subset (d) (404 d.o.f.).
}
\end{figure}

The velocity distribution parameters determined for each subset were used to fit the corresponding spectra with the theoretical lineshape.
A fifth subset of spectra was recorded in 2013 at higher pressure values: they were fitted using the velocity distribution parameters of the subset (a), while including in quadrature an additional uncertainty, corresponding to the variability between the four velocity distribution determinations.
This added uncertainty entails a correlation between the 2013 and the 2016-2017 data sets, which corresponds to a covariance of 2.6 kHz$^2$.

All our frequency measurements are performed with reference to a hydrogen maser, at LNE-SYRTE, which shows a relative daily drift of the order of $10^{-16}$ as compared to the Cs clock realizing the SI second \cite{syrte}. This drift was modeled and corrected for each subset of spectra. 

Figure \ref{Fig_spectres} shows the average of the spectra from subset (c) at two different magnetic fields, fitted by our theoretical lineshape.
No variability of the experimental lineshape, and notably of its broadening, appears when comparing spectra to their average.
Besides, examination of the residuals reveals no identifiable pattern, over all magnetic fields and subsets of spectra.
This supports the use of our theoretical lineshape, first to fit the parameters of the theoretical velocity distribution onto the velocity-dependent $B$-induced frequency shift shown on Figure \ref{dispersion}; and second to fit all recorded spectra and extract values of $\nu_{1S-3S}$, as illustrated Figure \ref{allspectra}.

Performing a separate analysis for the aforementioned subsets of spectra allows us to assess, within our uncertainties, the consistency of our outcome and the absence of noticeable drift of our measurements. That these subsets of spectra were recorded at different hydrogen pressure values moreover substantiates the choice of a pressure-independent theoretical velocity distribution (Figure \ref{distribvitesses}).

\subsection{Light and pressure shifts}
\begin{figure}
  \includegraphics[width=\columnwidth]{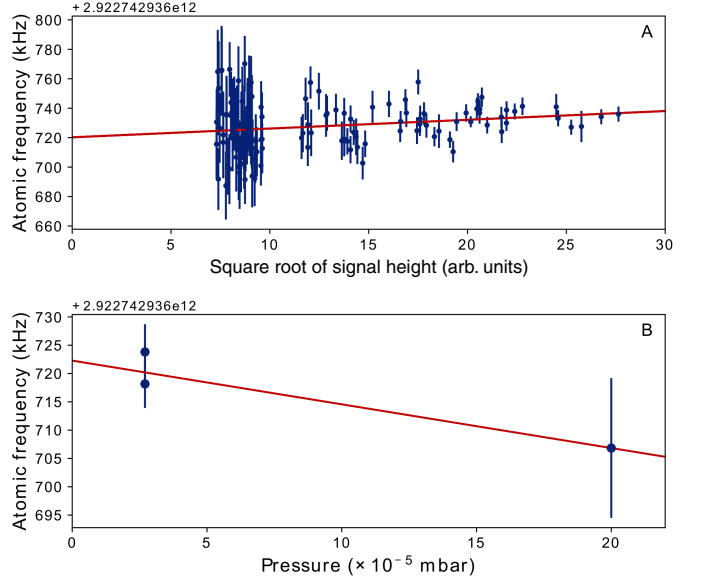}%
  \caption{\label{extrapolations}%\col 
%%% \col is to be used here for color figures (only).
    Experimental extrapolations performed on the 2016-2017 data to determine the light shift (A) and pressure shift (B) corrections.}
\end{figure}

The light shift (LS) and pressure shift (PS) corrections were evaluated independently for the 2013 and the 2016-2017 campaigns. As an illustration, Table 1 summerizes the average corrections applied for each of the four subsets of spectra previously mentioned.
Figure \ref{extrapolations} presents the extrapolation achieved for the 2016-2017 campaign.

First, considering a set of spectra obtained at a given pressure, a linear regression was performed on the values $\nu_{1S-3S}$ as a function of an indicator of the light intensity inside the build-up cavity. 
Two such indicators could be used: the 
potential difference
of the photodiode monitoring the UV power transmitted by the build-up cavity, and the square root of the fluorescence signal height at resonance. 

The former has the drawback of being more sensitive to the realignment of the optical cavity.
Besides, the photodiode tends to be damaged by the UV light, and it can be required to interface it by a fluorescent medium (in our case, a fluorescein solution).
The latter needs to be corrected at non-zero magnetic field, since certain sub-transitions can contribute less to the signal, being Zeeman shifted or Stark broadened. 
It must also be reevaluated for data sets corresponding to different hydrogen pressure.

As both indicators yielded congruent results, the most precise one was retained for each data set. 
For the 2013 campaign for instance, the average light shift correction was determined to be $\delta_{LS}=-5.9(1.2)$ kHz using the transmitted power, and $-5.9(1.6)$ kHz using the square root of the signal height. 

Once corrected from the light shift, all spectra corresponding to a given hydrogen pressure were then averaged, and a linear regression was performed with respect to the pressure. 
The hydrogen pressure was monitored by an ionization gauge placed aside the interaction region, only providing a relative measurement. 
As the gauge was replaced in 2014, no precise comparison could be done between the pressure measurements performed in 2013 and those performed in 2016-2017.
Without hydrogen, the background pressure in the vacuum chamber was $2\times 10^{-6}$ mbar. 

%\begin{table}
%  \begin{andptabular}[]{l|c|c|c|c}{Average frequency corrections for each subset of spectra, in kHz. The apparent frequency $\nu_{app}$, as obtained from the simple Lorentzian fit of the experimental spectra, and the final frequency $\nu_{1S-3S}^{F=1}$, are given subtracted from $2\ 922\ 742\ 936\ 000$ kHz.}
%Data set & a & b & c & d\\
%\hline
%$\nu_{app}$ &592.2(0.7)& 596.8(0.9)&594.4(1.1)&581.6(2.2) \\
%\hline
%$+\delta_{SOD}$    &133.2(1.3)& 138.0(3.8) &136.5(2.1) &132.2(6.8) \\
%$+\delta_{LS}$&-5.9(1.2)&-10.4(3.0)&-12.1(3.6)&-6.3(10.2) \\
%$+\delta_{PS}$ &3.6(2.0)& \multicolumn{3}{c}{Pressure extrapolation} \\
%$+\delta_{Maser}$ & -0.599(6)&   \multicolumn{3}{c}{-1.043(6)}       \\
%\hline
%$=\nu_{1S-3S}^{F=1}$&722.5(2.8)& \multicolumn{3}{c}{721.9(4.9)} \\
%  \end{andptabular}
%\end{table}

\subsection{Investigation on systematic effects}
Several other systematic effects have been investigated, that appeared to be negligible. Stray electromagnetic fields, in particular, were considered.
Compensation coils were placed around the atom-laser interaction chamber.
Moreover, the absence of a noticeable residual Zeeman effect was evaluated by frequently reversing the current direction in the coils producing the transverse $\vec{B}$ field, and by verifying that no shift was thus induced.  

A possible residual Stark effect was also studied.
The amplitude of the stray electric field that would be required to shift by at most 400 Hz the apparent transition frequency is 10 mV/cm \cite{theseSandrine}.
Particular care has been taken to avoid such an electric field in the atom-laser interaction chamber, covered with aquadag paint.
Furthermore, for the considered energy levels of hydrogen, blackbody radiation shifts are negligible \cite{farley}.

Other effects were examined, that could have affected the velocity distribution determination. 
Notably, the gaussian geometry of the laser beam in the Fabry-Perot cavity causes 
the probability of detection of a fluorescence photon to depend on the velocity of the atom that emitted it.
Indeed, faster atoms can de-excite in the detection region while having been excited further away from it, where the laser intensity is weaker.
This effect has been simulated and included in our theoretical lineshape; it amounts to less than 1 Hz \cite{olivthese}.

Eventually, a background noise is detected by the photomultiplier: mostly due to UV-induced fluorescence within the detection chamber and optics, it shows no dependency with respect to the UV frequency. Scanned over a large frequency range of 11 MHz, the transition exhibits a flat tail; when fitted with the theoretical lineshape, it yields the very same result than scanned over a smaller range.

It is also worth noting that, despite the changes in the experimental setup operated in between, the 2013 and the 2016-2017 measurement campaigns are in very good agreement.
This seems to suggest that the frequency measurement scheme, in particular, does not conceal under-estimated systematic effects. Indeed, the frequency comb was back then replaced, the purposedly redundant frequency counting setup was rearranged, the transfer laser between 1064 and 532 nm was installed, and the double-pass AOM for phase stabilization was implemented.

\subsection{Results}
The results of the 2013 and the 2016-2017 campaigns agree to within 1 kHz. 
We calculate their weighted average, caracterizing their correlation with the covariance mentioned above. 
In order to obtain the centroid frequency of the $1S-3S$ transition, a hyperfine correction of $+341\ 949.077(3)$ kHz is applied, derived from measurements of the $1S$ and $2S$ hyperfine splittings \cite{karshenboim}. 
The resulting value is:
\begin{equation}
\nu_{1S-3S}=2\ 922\ 743\ 278\ 671.5(2.6)\text{ kHz}
\end{equation}
The values of the Rydberg constant and the proton radius that can be derived by combining this result with the $1S-2S$ transition frequency \cite{parthey} are:
\begin{equation} \begin{array}{l}
R_\infty = 10\ 973\ 731.568\ 53(14)\text{ m}^{-1} \\
r_p = 0.877(13)\text{ fm} 
\end{array} \end{equation}
As already mentioned, they are in very good agreement with the CODATA-2014 values, and disagree by $2.8\sigma$ with those deduced from muonic hydrogen spectroscopy. 
Although not statistically significant, this disagreement impels us to continue  investigating possible systematic effects.

\section{Ongoing work} 
\begin{figure}
  \includegraphics[width=\columnwidth]{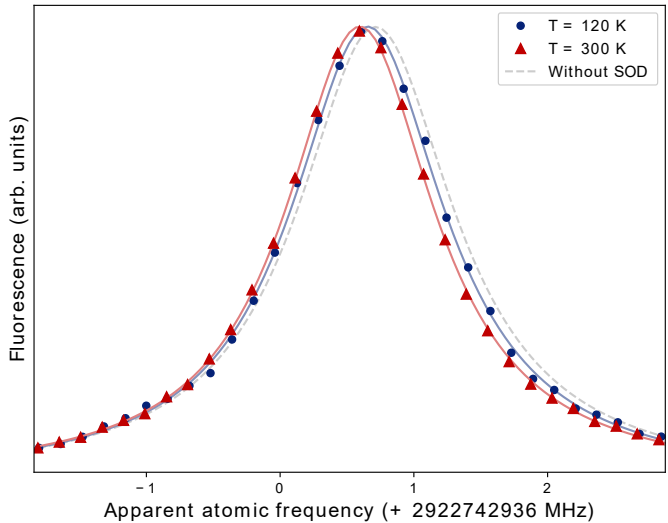}%
  \caption{\label{spectres}%\col 
Spectra of the $1S-3S$ transition in hydrogen, uncorrected from systematic effects, obtained with a nozzle at room temperature and at 120 K. The experimental datapoints, each corresponding to about eight minutes of integration time, are fitted by a Lorentzian lineshape.
The dashed line is a simulation of the line position at 0 K.}
\end{figure}

Our current efforts aim at remeasuring the $1S-3S$ transition frequency once more, under different experimental conditions.
More precisely, in order to both reduce our main systematic effect and cross-check the SOD correction method with a different velocity distribution, we are currently proceeding to the cooling of our atomic beam. By passing through an Aluminum-made nozzle cooled down by liquid nitrogen, the hydrogen atoms experience a SOD shift reduced by 50 to 60$\%$. Preliminary results are shown on Figure \ref{spectres}. 

Furthermore, our experimental setup allows for the spectroscopy of the $1S-3S$ transition in deuterium. As a heavier atom, deuterium has the advantage of being less sensitive to the SOD effect than hydrogen. 
The frequency metrology of this transition, which has never been done, is all the more interesting as the spectroscopy of muonic deuterium also revealed a discrepancy among the determinations of the deuteron charge radius \cite{deuterium}.
We first observed this transition in 2016; further measurements are underway.

Thereafter, in order to investigate possible systematic effects related to the configuration of our hydrogen beam, an entirely new effusive beam is to be build, pumped by an oil-free vacuum system.

\section*{Conclusion}
The $1S-3S$ transition frequency of the hydrogen atom has been measured in our group with a relative uncertainty of $9\times 10^{-13}$. 
This result, combined with the $1S-2S$ transition frequency, yields values of the Rydberg constant and the proton charge radius that are in good agreement with the current CODATA-recommended values. 
These latter, however, disagree with other recent results, from both hydrogen and muonic hydrogen spectroscopy.
Investigations are therefore ongoing to understand possible sources of this disagreement.

\subsection*{Acknowledgments}
The authors thank O. Acef for the transfer laser.
This work was supported by the French National Research Agency (ANR) through
the cluster of excellence FIRST-TF (ANR-10-LABX-48), the PROCADIS project 
(ANR-2010-BLANC:04510) and the Equipex REFIMEVE+ (ANR-11-EQPX-0039), and by the 
CNRS.


\begin{thebibliography}{00}

%\bibitem{bib.aldoscience} A. Antognini, F. Nez, K. Schuhmann \emph{et al}, ``Proton structure from the measurement of 2S-2P transition frequencies of muonic hydrogen", \emph{Science}, vol. 339, no. 6118, pp. 417-420, 2013.

\bibitem{nature2010}
\textsc{R.~Pohl},
\textsc{A.~Antognini},
\textsc{F.~Nez} \emph{et~al.}, 
Nature \textbf{466}, 213-216 (2010).

\bibitem{aldoscience}%
\textsc{A.~Antognini}, 
\textsc{F.~Nez},
\textsc{K.~Schuhmann}
\emph{et~al.}, 
%``Proton structure from the measurement of 2S-2P transition frequencies of muonic hydrogen", 
Science \textbf{339}, 6118 (2013).

\bibitem{codata2014} 
\textsc{P.\,J.~Mohr}, 
\textsc{D.\,B.~Newell} and 
\textsc{B.\,N.~Taylor}, %``CODATA recommended values of the fundamental physical constants: 2014", \emph{Rev. Mod. Phys. J. D}, vol. 88, p. 035009, 2016.
Rev. Mod. Phys. J. D \textbf{88}, 035009 (2016).

\bibitem{carlson2015}
\textsc{C.\,E.~Carlson},
Prog. Part. Nucl. Phys. \textbf{82}, 59-77 (2015).

\bibitem{beyer2017} 
\textsc{A.~Beyer},
\textsc{L.~Maisenbacher},
\textsc{A.~Matveev} \emph{et~al.}, % \emph{et al.}, ``The Rydberg constant and proton size from atomic hydrogen", \emph{Science}, vol. 358, p. 79, 2017.
Science \textbf{358}, 79 (2017).

\bibitem{fleurbaey18}
\textsc{H.~Fleurbaey},
\textsc{S.~Galtier},
\textsc{S.~Thomas} \emph{et~al.},
Phys. Rev. Lett. \textbf{120}, 183001 (2018).

\bibitem{parthey}
\textsc{C.\,G.~Parthey},
\textsc{A.~Matveev},
\textsc{J.~Alnis} \emph{et~al.},
Phys. Rev. Lett. \textbf{107}, 203001 (2011).

\bibitem{bourzeix96}
\textsc{S.~Bourzeix},
\textsc{B.~de~Beauvoir},
\textsc{F.~Nez}  \emph{et~al.},
Phys. Rev. Lett. \textbf{76}, 384 (1996).

\bibitem{hagel2002}
\textsc{G.~Hagel},
\textsc{F.~Nez} and
\textsc{F.~Biraben},
Appl. Opt. \textbf{41}, 7702-7706 (2002).

\bibitem{sandrineoptcom} 
\textsc{S.~Galtier}, 
\textsc{F.~Nez}, 
\textsc{L.~Julien} and 
\textsc{F.~Biraben}, %``Ultraviolet continuous-wave laser source at 205 nm for hydrogen spectroscopy", \emph{Optics Comm.}, vol. 324, p. 34-37, 2014.
Optics Comm. \textbf{324}, 34-37 (2014).

\bibitem{rubidium}
\textsc{L.~Hilico},
\textsc{R. Felder}, 
\textsc{D. Touahri}, 
\textsc{O. Acef}, 
\textsc{A. Clairon} and
\textsc{F. Biraben}, 
Eur. Phys. J. AP \textbf{4}, 219 (1998).

\bibitem{epjd1998}
\textsc{B.~de~Beauvoir},
\textsc{F.~Nez},
\textsc{L.~Hilico} \emph{et~al.},
Eur. Phys. J. D \textbf{1}, 227 (1998).

\bibitem{gaetPRL} 
\textsc{G.~Hagel}, 
\textsc{R.~Battesti}, 
\textsc{F.~Nez}, 
\textsc{L.~Julien} and 
\textsc{F.~Biraben}, %``Observation of a motional Stark effect to determine the second-order Doppler effect",
Phys. Rev. Lett. \textbf{89}, 203001 (2002).

\bibitem{biraben91} 
\textsc{F.~Biraben}, 
\textsc{L.~Julien}, 
\textsc{J.~Plon} and 
\textsc{F.~Nez}, 
%``Compensation of the second-order Doppler effect in two-photon spectroscopy of atomic hydrogen", 
Europhys. Lett. \textbf{15}, 8  (1991). % vol. 15, no. 8, pp. 831-836, 1991.

\bibitem{gaetthese} 
\textsc{G.~Hagel}, 
Ph.D. thesis, https://hal.archives-ouvertes.fr/tel-00000848.

\bibitem{BetheSalpeter}
\textsc{H.\,A.~Bethe} and
\textsc{E.\,E.~Salpeter},
Quantum mechanics of one- and two-electron atoms 
(Springer-Verlag, Berlin, 1957), sect. 47.

\bibitem{biraben73}
\textsc{B.~Cagnac},
\textsc{G.~Grynberg} and
\textsc{F.~Biraben},
J. Phys. France \textbf{34}, 845 (1973).

\bibitem{arnoult2010} 
\textsc{O.~Arnoult}, 
\textsc{F.~Nez}, 
\textsc{L.~Julien} and 
\textsc{F.~Biraben}, % ``Optical frequency measurement of the 1S-3S two-photon transition in hydrogen", \emph{Eur. Phys. J. D}, vol. 60, pp. 243-256, 2010.
Eur. Phys. J. D \textbf{60}, 243-256 (2010).

\bibitem{biraben79}
\textsc{F.~Biraben},
\textsc{M.~Bassini} and
\textsc{B.~Cagnac},
J. Phys. France \textbf{40}, 445 (1979).

\bibitem{hessels10}
\textsc{M.~Horbatsch} and 
\textsc{E.\,A.~Hessels}, 
Phys. Rev. A \textbf{82}, 052519 (2010).

\bibitem{hessels11}
\textsc{M.~Horbatsch} and 
\textsc{E.\,A.~Hessels}, 
Phys. Rev. A \textbf{84}, 032508 (2011).

\bibitem{fleurbaey2017} 
\textsc{H.~Fleurbaey}, 
\textsc{F.~Biraben}, 
\textsc{L.~Julien}, 
\textsc{J.-P.~Karr} and 
\textsc{F.~Nez}, %``Cross-damping effects in $1S-3S$ spectroscopy of hydrogen and deuterium", \emph{Phys. Rev. A}, vol. 95, p. 052503, 2017.
Phys. Rev. A \textbf{95}, 052503 (2017).

\bibitem{JPCRD15} 
\textsc{S.~Galtier}, 
\textsc{H.~Fleurbaey}, 
\textsc{S.~Thomas},% \emph{et~al.},
\textsc{L.~Julien}, 
\textsc{F.~Biraben} and 
\textsc{F.~Nez}, 
%``Progress in spectroscopy of the 1S-3S transition in hydrogen", \emph{J. Phys. Chem. Ref. Data}, vol. 44, p. 031201, 2015.
J. Phys. Chem. Ref. Data \textbf{44}, 031201 (2015).

\bibitem{olander} 
\textsc{D. R. Olander}, 
\textsc{R. H. Jones} and 
\textsc{W. J. Siekhaus}, 
J. App. Phys. \textbf{41}, 4388-4391 (1970).

\bibitem{champenois}
\textsc{C.~Champenois},
\textsc{M.~Jacquey},
\textsc{S.~Lepoutre} \emph{et~al.},
Phys. Rev. A \textbf{77}, 013621 (2008).

\bibitem{theseHelene} 
\textsc{H.~Fleurbaey}, 
Ph.D. thesis, https://hal.archives-ouvertes.fr/tel-01633631. 

\bibitem{syrte}
\textsc{G.\,D.~Rovera},
\textsc{S.~Bize}, 
\textsc{B.~Chupin},
\textsc{J.~Gu\'e na},
\textsc{P.~Laurent} and
\textsc{P.~Rosenbusch},
Metrologia \textbf{53}, S81 (2016).

\bibitem{theseSandrine} 
\textsc{S.~Galtier}, 
Ph.D. thesis, https://hal.archives-ouvertes.fr/tel-01080669.

\bibitem{farley}
\textsc{J.\,W.~Farley} and
\textsc{W.\,H~Wing},
Phys. Rev. A \textbf{23}, 2397 (1981).

\bibitem{olivthese} 
\textsc{O.~Arnoult}, 
Ph.D. thesis, https://hal.archives-ouvertes.fr/tel-00441568.

\bibitem{karshenboim}
\textsc{S.\,G.~Karshenboim} and
\textsc{V.\,G.~Ivanov},
Eur. Phys. J. D \textbf{19}, 13 (2002).

\bibitem{deuterium}
\textsc{R.~Pohl},
\textsc{F.~Nez},
\textsc{L.\,M.\,P.~Fernandes} \emph{et~al.},
Science \textbf{353}, 669-673 (2016).

\end{thebibliography}
\end{document}